\documentstyle[aps,amssymb]{revtex}

\begin{document}
\title{PHASE TRANSITION IN ANYON SUPERCONDUCTIVITY AT FINITE TEMPERATURE}
\author{{\it E. J. Ferrer\thanks{%
ferrer@fredonia.edu} and V. de la Incera\thanks{%
incera@fredonia.edu}}}
\author{{\it Dept. of Physics, State University of New York, Fredonia, NY 14063, USA}}
\date{SUNY-FRE-98-04}
\maketitle

\begin{abstract}
The magnetic response of the charged anyon fluid at temperatures larger than
the fermion enery gap ($T\gg \omega _{c}$) is investigated in the
self-consistent field approximation. In this temperature region a new phase,
characterized by an inhomogeneous magnetic penetration, is found. The
inhomogeneity is linked to the existence of an imaginary magnetic mass which
increases with the temperature at $T\gg \omega _{c}$. The system stability
in the ($T\gg \omega _{c}$)-phase is proved by investigating the
electromagnetic field rest-energy spectrum.
\end{abstract}

\section{Introduction}

In recent years there has been much interest in investigating
(2+1)-dimensional gauge theories with Chern-Simons (CS) interactions. This
interest is due, in part, to a variety of significant physical applications
of these theories in QFT, as well as in condensed matter physics.

A well known example in QFT is the Deser, Jackiw and Templeton Parity Anomaly%
\cite{Deser}. This result shows that the fluctuations of a massive Fermi
field induces a CS term in the effective action of the gauge fields. In this
context the photon acquires a topological mass\cite{Deser},\cite{Schonfeld}.
Parity violation\cite{Parity} and vortex solutions in lower dimensions\cite
{Vortex} are among the consequences of the presence of CS terms. On the
other hand, it has been shown that when Dirac fermions are coupled to a
Maxwell-Chern-Simons (MCS) gauge field, the Lorentz symmetry can be
spontaneously broken by the dynamical generation of a magnetic field \cite
{Lorentz}. The effects of CS terms in supersymmetric models have been also
investigated\cite{SuperS}.

In condensed matter, CS models have been considered in the study of
different physical applications. CS theories involving several vector
potentials are known to be particularly appropriate for describing the
quantum Hall effect\cite{Hall},\cite{Prange},\cite{Frad-Book}. A recent
model of the fractional quantum Hall effect considers that the electrons are
transformed into composite fermions by attaching two artificial statistical
flux quanta to each electron\cite{Jain}. The gauge field theory describing
these composite fermions introduces the statistical gauge flux via CS fields 
\cite{Lopez}.

When matter field is coupled to the CS gauge fields in (2+1)-dimensions, a
suitable description for {\it anyons} is obtained\cite{Hagen},\cite{Arova}.
Anyons \cite{Anyons}, \cite{Wilczek} are particles with fractional
statistics in (2+1)-dimensions. The anyon description within the CS gauge
theory is equivalent to attaching flux tubes to the charged fermions. The
Aharonov-Bohm phases resulting from the adiabatic transport of two anyons is
the source of the fractional exchange statistics\cite{Wilczek}.

It has been argued that strongly correlated electron systems in two
dimensions can be described by an effective field theory of anyons \cite{5}, 
\cite{5a}. Anyons can be also obtained as solitons which fractional spin in
electron systems. Excitations with fractional spin in two dimensional
systems necessarily obey fractional statistics \cite{5aa}. An important
feature of the anyon theory is that it violates parity and time-reversal
invariances. Although there are claims that anyons could play a basic role
in high-T$_{C}$ superconductivity \cite{5a}-\cite{6b}, at present no
experimental evidences of P and T violation in high-T$_{C}$
superconductivity have been confirmed. It should be pointed out,
nevertheless, that it is possible to construct more sophisticated P and T
invariant anyonic models\cite{6a}.

Whether linked or not to high-T$_{C}$ superconductivity, anyon
superconductivity is an interesting effect in its own right, and deserves a
deeper study. As it is known, anyon superconductivity has an origin
different from the Nambu-Goldstone-Higgs like mechanism. The genesis of the
anyon superconductivity is given by the spontaneously violation of
commutativity of translations in the free anyon system\cite{Chen}. This new
mechanism may have wider applications than the original physical problem
that motivated its study.

The superconducting behavior of anyon systems at $T=0$ has been investigated
by many authors \cite{Arova}, \cite{6}-\cite{15a}. At $T=0$, anyon
superconductivity appears due to the exact cancellation between the bare and
induced CS terms in the effective action of the theory\cite{Bank}. However,
at $T\neq 0$ this cancelation does not take place \cite{9}. Hence, several
authors \cite{9}-\cite{14} have advocated that the superconducting phase
evaporates at any finite temperature. In ref. \cite{8}, it has been
independently claimed that the destruction of the anyon superconducting
phase at finite temperature is connected to the existence of a long-range
mode, found inside the infinite bulk at $T\neq 0$. This long range mode is
the consequence of the existence of a pole $\sim \left( \frac{1}{{\bf k}^{2}}%
\right) $ in the polarization operator component $\Pi _{00}$ at finite
temperature. On the other hand, in Ref. \cite{Cabra} it has been argued that
an anyon fluid should undergo a Kosterlitz-Thouless type transition rather
than an immediate destruction of its superconducting state for all $T>0$.

In previous works \cite{Our} we found that, contrary to some authors'
belief, the superconducting behavior, manifested through the Meissner effect
in the charged anyon fluid at $T=0$, does not disappear as soon as the
system is heated. In papers \cite{Our} we showed that the presence of
boundaries affects the dynamics of the two-dimensional system in such a way
that the long-range mode, that accounts for a homogeneous field penetration 
\cite{8}, cannot propagate in the bulk. According to these results, the
anyon system with boundaries exhibits a total Meissner effect at
temperatures smaller than the fermion energy gap ($T\ll \omega _{c}$). In
this case the magnetic field cannot penetrate the bulk farther than a very
short distance ($\overline{\lambda }\sim 10^{-5}cm$ for $T\sim 200$ $K$ and
electron densities characteristic of the high -T$_{c}$ superconductors). Our
main conclusion was that the magnetic behavior of the anyon fluid is not
just determined by its bulk properties, but it is essentially affected by
the sample boundary conditions. The importance of the boundary conditions in
(2+1)-dimensional models has been previously stressed in ref.\cite{b}.

It is natural to expect that at temperatures larger than the energy gap this
superconducting behavior should not exist. At those temperatures the
electron thermal fluctuations should make accessible the free states
existing beyond the energy gap. As a consequence, the charged anyon fluid
should not be a perfect conductor at $T\gg \omega _{c}$. A signal of such a
transition may be found studying the system magnetic response at those
temperatures. The main goal in this paper is to investigate the
characteristics of the magnetic response in this high temperature phase.

In what follows we show that at $T\gg \omega _{c}$ an externally applied
constant and homogeneous magnetic field can penetrate the sample, giving
rise to a periodic inhomogeneous magnetic field within the bulk. The
inhomogeneity of the magnetic response increases with the temperature. We
also find that, contrary to the $T\ll \omega _{c}$ case, a long-range
penetration, associated to a massless mode of the electromagnetic field
within the anyon fluid, can penetrate the bounded sample in the high
temperature phase. Nevertheless, in the range of temperatures considered the
effect of this homogeneous field penetration is negligible as compared to
the inhomogeneous component.

These results corroborate the existence of a phase transition in the charged
anyon fluid from a superconducting phase, at $T\ll \omega _{c}$ (see Ref. 
\cite{Our}), to a non-superconducting phase, at $T\gg \omega _{c}$. The
inhomogeneous character of the magnetic response in the high temperature
phase is linked to the inhomogeneity of the spatial distribution of the
induced many-particle charge and current densities at $T\gg \omega _{c}$.

We also prove that the inhomogeneous magnetic penetration is associated to
an imaginary magnetic mass associated to one of the electromagnetic field
modes within the charged anyon fluid. The appearance of an imaginary
magnetic mass at $T\gg \omega _{c}$ by no means indicates that the linear
approximation used in these calculations is broken by the presence of
tachyons. Tachyons, as it is known, correspond to imaginary rest-energy
solutions. The rest energy and the magnetic mass spectrum of the MCS theory
at finite density are not the same at $T\gg \omega _{c}$, as proved in Sec.
4. For the MCS theory at finite density, we show that the rest energies of
the electromagnetic field modes at $T\gg \omega _{c}$ are real.

The plan for the paper is as follows. In Sec. 2, for completeness, as well
as for the convenience of the reader, we review the many-particle
(2+1)-dimensional MCS model used to describe the charged anyon fluid in the
self-consistent field approximation. Principal attention is given to the
derivation of the high-temperature MCS effective action in the linear
approximation. In Sec. 3 we study the magnetic response in the
self-consistent field approximation of a charged anyon fluid confined to a
half plane in the ($T\gg \omega _{c}$)-phase. We find the analytical
solution of the MCS field equations that satisfies the corresponding
boundary conditions and minimizes the system free energy. The propagation
modes of the magnetic field within the anyon fluid have three contributions:
one that decays exponentially in space, other that accounts for an
homogeneous field penetration, and a third that changes periodically in
space within the sample. When all the coefficients appearing in the magnetic
field solution are evaluated for the range of temperatures and other
characteristic parameter values, we find that the leading term in the
magnetic response is an inhomogeneous magnetic field with a characteristic
wavelength that decreases with the temperature. In Sec. 4 we investigate the
dispersion equation for the Maxwell field in the high-temperature
approximation. We solve this equation in arbitrary covariant gauges for the
Maxwell and CS fields to obtain the magnetic masses of the charged anyon
fluid in the high-temperature phase. As expected, the magnetic masses turn
out to be equal to the inverse length scales which characterize the magnetic
response of the anyon fluid in this phase. We then prove that the existence
of an imaginary magnetic mass cannot be associated to a tachyonic mode in
this many-particle system with CS interactions. In doing that, we find the
rest energies of the electromagnetic field modes at $T\gg \omega _{c}$. Sec.
6 contains the summary and discussion.

\section{Anyon fluid at high temperature and density}

\subsection{Many-particle model and energy gap}

The non-relativistic charged anyon system in interaction with an
electromagnetic field in 2+1 dimensions can be modeled by the MCS Lagrangian
density

\begin{equation}
{\cal L}=-\frac{1}{4}F_{\mu \nu }^{2}-\frac{N}{4\pi }\varepsilon ^{\mu \nu
\rho }a_{\mu }\partial _{\nu }a_{\rho }+en_{e}A_{0}+i\psi ^{\dagger
}D_{0}\psi -\frac{1}{2m}\left| D_{k}\psi \right| ^{2}.  \eqnum{2.1}
\end{equation}
In (2.1) $\mu ,\nu ,\rho =0,1,2$ and $k=1,2.$ $A_{\mu }$ and $a_{\mu }$
represent the electromagnetic field and the CS field respectively. The CS
fields are simply changing the quantum statistics of the matter field, thus,
they do not have independent dynamics. $\psi $ represents non-relativistic
spinless fermions. $N\ $ is a positive integer which determines the
magnitude of the CS coupling constant. $n_{e}$ is a background neutralizing
`classical' charge density, and $m$ is the fermion mass. The covariant
derivative $D_{\nu }$ (we use the metric $g_{\mu \nu }$=$(1,-\overrightarrow{%
1})$) is given by

\begin{equation}
D_{\nu }=\partial _{\nu }+i\left( a_{\nu }+eA_{\nu }\right) ,\qquad \nu
=0,1,2  \eqnum{2.2}
\end{equation}

The many-particle system is implemented through the Grand Partition function

\begin{equation}
{\cal Z}=Tr\exp \left[ -\beta {\cal H}\left( \mu \right) \right]  \eqnum{2.3}
\end{equation}
where $\beta $ is the inverse of the absolute temperature and the
many-particle Hamiltonian density is given by

\begin{equation}
{\cal H}\left( \mu \right) ={\cal H}-\mu {\cal N}  \eqnum{2.4}
\end{equation}
with ${\cal H}$ the canonical Hamiltonian density corresponding to the
Lagrangian density (2.1) and $\mu $ the chemical potential associated to the
conserved particle number ${\cal N}=\psi ^{\dagger }\psi $.

Doing the functional integrals in the momenta in (2.3) we obtain the
effective-Lagrangian density of the many-particle anyon system

\begin{equation}
{\cal L}_{eff}={\cal L}+\psi ^{\dagger }\mu \psi  \eqnum{2.5}
\end{equation}

The mean-field Euler-Lagrange equations derived from (2.5) are

\begin{mathletters}
\begin{equation}
-\frac{N}{4\pi }\varepsilon ^{\mu \nu \rho }f_{\nu \rho }=\left\langle
j^{\mu }\right\rangle  \eqnum{2.6}
\end{equation}

\begin{equation}
\partial _{\nu }F^{\mu \nu }=e\left\langle j^{\mu }\right\rangle
-en_{e}\delta ^{\mu 0}  \eqnum{2.7}
\end{equation}
To guarantee the electric neutrality of the system formed by the electron
fluid and the background charge $n_{e}$ we impose the condition

\end{mathletters}
\begin{equation}
\left\langle j^{0}\right\rangle -n_{e}\delta ^{\mu 0}=0,  \eqnum{2.8}
\end{equation}
where $\left\langle j^{0}\right\rangle $ is the many-particle system fermion
density 
\begin{equation}
\left\langle j^{0}\right\rangle =\frac{\partial \Omega }{\partial \mu }, 
\eqnum{2.9}
\end{equation}
with $\Omega $ is the one-loop fermion thermodynamic potential

\begin{equation}
\Omega =-\beta ^{-1}\ln \det G_{0}^{-1}\left( p,\mu \right) ,  \eqnum{2.10}
\end{equation}
given in terms of the free fermion thermal Green's function

\begin{equation}
G_{0}\left( p,\mu \right) =\frac{1}{ip_{4}+\mu -\frac{{\bf p}^{2}}{2m}} 
\eqnum{2.11}
\end{equation}
where $p_{4}=(2n+1)\pi /\beta $, are the discrete frequencies $(n=0,1,2,...)$
corresponding to fermion fields.

The existence of a different from zero fermion density, as it is required by
the neutrality condition (2.8), generates through eqs.(2.6)-(2.7) a
nontrivial background of CS magnetic field

\begin{equation}
\overline{b}=\overline{f}_{21}=\frac{2\pi n_{e}}{N}.  \eqnum{2.12}
\end{equation}

Then, the unperturbed one-particle Hamiltonian of the matter field
represents, in this many-particle system, a particle in the background of
that CS magnetic field $\overline{b\text{,}}$

\begin{equation}
H_{0}=-\frac{1}{2m}\left[ \left( \partial _{1}+i\overline{a}_{1}\right)
^{2}+\partial _{2}^{2}\right]  \eqnum{2.13}
\end{equation}
In (2.13) we considered the background CS potential, $\overline{a}_{k}$, $%
(k=1,2)$, in the Landau gauge

\begin{equation}
\overline{a}_{k}=\overline{b}x_{2}\delta _{k1}  \eqnum{2.14}
\end{equation}

The eigenvalue problem defined by the Hamiltonian (2.13) with periodic
boundary conditions in the $x_{1}$-direction: $\Psi \left( x_{1}+L,\text{ }%
x_{2}\right) =$ $\Psi \left( x_{1},\text{ }x_{2}\right) $,

\begin{equation}
H_{0}\Psi _{nk}=\epsilon _{n}\Psi _{nk},\qquad n=0,1,2,...\text{ }and\text{ }%
k\in {\cal Z}  \eqnum{2.15}
\end{equation}
has eigenvalues and eigenfunctions given respectively by

\begin{equation}
\epsilon _{n}=\left( n+\frac{1}{2}\right) \omega _{c}\qquad  \eqnum{2.16}
\end{equation}

\begin{equation}
\Psi _{nk}=\frac{\overline{b}^{1/4}}{\sqrt{L}}\exp \left( -2\pi
ikx_{1}/L\right) \varphi _{n}\left( x_{2}\sqrt{\overline{b}}-\frac{2\pi k}{L%
\sqrt{\overline{b}}}\right)  \eqnum{2.17}
\end{equation}
where $\omega _{c}=\overline{b}/m$ is the cyclotron frequency and $\varphi
_{n}\left( \xi \right) $ are the orthonormalized harmonic oscillator wave
functions.

Note that the energy levels $\epsilon _{n}$ are degenerates (they do not
depend on $k$). Therefore, for each Landau level $n$, there exists a band of
degenerate states. The cyclotron frequency $\omega _{c}$ plays here the role
of the energy gap between occupied Landau levels. It is easy to prove\cite
{Prange} that the number of states per unit area of a full Landau level is
proportional to the background CS magnetic field $\left( n_{B}=\overline{b}%
/2\pi \right) $. Thus, the filling factor, defined as the ratio between the
density of particles $n_{e}$ and the number of states per unit area of a
full Landau level $n_{B}$, is equal to the CS coupling constant $N$.

When the magnitude of the CS coupling constant $N$ is considered as a
positive integer, the system will have $N$ completely filled Landau levels.
Once this ground state is established, it can be argued immediately \cite{6}%
, \cite{6b}, \cite{Chen}, \cite{15}, that at $T=0$ the system will be
confined to a filled band, which is separated by an energy gap ( $\omega
_{c} $) from the free states; therefore, it is natural to expect that at $%
T=0 $ the system should superconduct. This result is already a well
established fact on the basis of Hartree-Fock analysis\cite{6} and Random
Phase Approximation \cite{6b},\cite{Chen}. At $T\neq 0$ it was proved in
Refs. \cite{Our} that the existence of a natural scale (the cyclotron
frequency $\omega _{c}$) in this theory, makes possible the realization of a
superconducting phase at $T\ll \omega _{c}$, for systems confined to a
bounded region.

It is logical to expect that this energy scale, $\omega _{c}$, separates two
different physical phases of the system. As we will prove below, the
superconducting state, found at $T\ll \omega _{c}$, disappears when the
system reaches temperatures large enough (i.e. at $T\gg \omega _{c}$) to
move the electrons beyond the energy gap to the free-energy band.

\subsection{Effective action in the linear approximation}

The linear response of the medium can be found under the assumption that the
quantum fluctuations of the gauge fields about the ground-state are small.
In this case the one-loop fermion contribution to the effective action,
obtained after integrating out the fermion fields, can be evaluated up to
second order in the gauge fields. The effective action in terms of the
quantum fluctuation of the gauge fields within the linear approximation \cite
{8},\cite{11} takes the form

\begin{equation}
\Gamma _{eff}\,\left( A_{\nu },a_{\nu }\right) =\int dx\left( -\frac{1}{4}%
F_{\mu \nu }^{2}-\frac{N}{4\pi }\varepsilon ^{\mu \nu \rho }a_{\mu }\partial
_{\nu }a_{\rho }\right) +\Gamma ^{\left( 2\right) }  \eqnum{2.18}
\end{equation}
$\Gamma ^{\left( 2\right) }$ is the one-loop fermion contribution to the
effective action in the linear approximation

\begin{equation}
\Gamma ^{\left( 2\right) }=\int dxdy\left[ a_{\mu }\left( x\right) +eA_{\mu
}\left( x\right) \right] \Pi ^{\mu \nu }\left( x,y\right) \left[ a_{\nu
}\left( y\right) +eA_{\nu }\left( y\right) \right] .  \eqnum{2.19}
\end{equation}

In (2.19) $\Pi _{\mu \nu }$ represents the fermion one-loop polarization
operator in the presence of the CS background magnetic field $\overline{b}$.
To calculate the different components of $\Pi _{\mu \nu }$, it is convenient
to consider its structure,

\begin{eqnarray}
\Pi ^{\mu \nu } &=&{\cal A}_{1}\left( k^{2}g^{\mu \nu }-k^{\mu }k^{\nu
}\right) +i{\cal A}_{2}\varepsilon ^{\mu \nu \rho }k_{\rho }  \nonumber \\
&&+{\cal A}_{3}\left( \delta _{ij}{\bf k}^{2}-k_{i}k_{j}\right) \delta ^{\mu
i}\delta ^{\nu j}  \eqnum{2.20}
\end{eqnarray}

The $\Pi _{\mu \nu }$ operator (2.20) is a second rank tensor, formed by the
momentum space basic tensors $k_{\nu }$, $g_{\mu \nu }$ and $\varepsilon
^{\mu \nu \rho }$. The structure of $\Pi _{\mu \nu }$ appearing in (2.20) is
chosen in such a way that it satisfies the symmetry properties of the
theory. That is, the transversality condition $\left( k_{\mu }\Pi _{\mu \nu
}=0\right) $, which is a consequence of the gauge invariance of the theory;
the rotational invariance in the two-dimensional space, and the invariance
under simultaneous permutation of indices and arguments $\left( \Pi _{\nu
\mu }\left( k\right) =\Pi _{\mu \nu }\left( -k\right) \right) $.

When the system has a rotational symmetry in the 2+1 space (as it is the
case of a relativistic invariant system \cite{Fradkyn}), the polarization
operator can be expressed in terms of only one independent invariant
coefficient (${\cal A}_{1}$ in eq.(2.20)). In the non-relativistic case, the
symmetry between time and space-components is broken, giving rise to an
additional invariant coefficient (${\cal A}_{3}$ in eq.(2.20)). In the anyon
fluid, in particular, as the P and T-invariances are broken, a third
independent coefficient arises (${\cal A}_{2}$ in eq.(2.20)). The presence
of a medium (which is the case when we have a statistical system), however,
does not introduce here any additional independent coefficient\cite{Efrain},
as it is the case in the relativistic formulation\cite{Fradkyn}. The reason
is that a medium cannot produce any new symmetry breaking in the
non-relativistic case, where the Minkowskian symmetry is already broken.

For the sake of convenience we write the independent coefficients: ${\cal A}%
_{1}$, ${\cal A}_{2}$ and ${\cal A}_{3}$, in term of the following new set ($%
{\it \Pi }_{{\it 0}}$, ${\it \Pi }_{{\it 0}}\,^{\prime }$, ${\it \Pi }_{{\it %
1}}$ and ${\it \Pi }_{\,{\it 2}}$) of coefficients

\begin{equation}
{\cal A}_{1}=\frac{{\it \Pi }_{{\it 0}}}{{\bf k}^{2}}+{\it \Pi }_{{\it 0}%
}\,^{\prime },\qquad {\cal A}_{2}={\it \Pi }_{{\it 1}},\qquad {\cal A}_{3}+%
{\cal A}_{1}={\it \Pi }_{\,{\it 2}}  \eqnum{2.21}
\end{equation}

From (2.20) and (2.21) we have that the new independent coefficients can be
found from the polarization operator components

\begin{mathletters}
\begin{equation}
\Pi ^{00}=\left( \frac{{\it \Pi }_{{\it 0}}}{{\bf k}^{2}}+{\it \Pi }_{{\it 0}%
}\,^{\prime }\right) \left( k^{2}-k_{0}^{2}\right)  \eqnum{2.22}
\end{equation}

\begin{equation}
\Pi ^{02}=-\left( \frac{{\it \Pi }_{{\it 0}}}{{\bf k}^{2}}+{\it \Pi }_{{\it 0%
}}\,^{\prime }\right) k^{0}k^{2}-i{\it \Pi }_{{\it 1}}k^{1}  \eqnum{2.23}
\end{equation}
\begin{equation}
\Pi ^{22}={\it \Pi }_{\,{\it 2}}k_{1}^{2}-\left( \frac{{\it \Pi }_{{\it 0}}}{%
{\bf k}^{2}}+{\it \Pi }_{{\it 0}}\,^{\prime }\right) k_{0}^{2}  \eqnum{2.24}
\end{equation}

Taking into account that we will investigate the magnetic response of the
charged anyon fluid to a uniform and constant applied magnetic field, we
need the $\Pi _{\mu \nu }$ leading behaviors for static $\left(
k_{0}=0\right) $ and slowly $\left( {\bf k}\sim 0\right) $ varying
configurations. Then, from (2.22)-(2.24) it is clear that to find the
independent coefficients ${\it \Pi }_{{\it 0}}$, ${\it \Pi }_{{\it 0}%
}\,^{\prime }$, ${\it \Pi }_{{\it 1}}$ and ${\it \Pi }_{\,{\it 2}}$ in this
limit we just need to calculate the $\Pi _{\mu \nu }$ components $\Pi _{00}$%
, $\Pi _{02}$ and $\Pi _{22}$ in the ($k_{0}=0,$ ${\bf k}\sim 0$)-limit.

The polarization operator components of the many-particle system are
calculated using the fermion thermal Green's function in the presence of the
background field $\overline{b}$ \cite{8}, \cite{11}, \cite{Efrain}

\end{mathletters}
\begin{eqnarray}
G\left( p_{4},{\bf p}\right) &=&\int\limits_{0}^{\infty }d\rho
\int\limits_{-\infty }^{\infty }dx_{2}\sqrt{\overline{b}}\exp -\left(
ip_{2}x_{2}\right) \exp -\left( ip_{4}+\mu -\frac{\overline{b}}{2m}\right)
\rho  \nonumber \\
&&\sum\limits_{n=0}^{\infty }\varphi _{n}\left( \xi \right) \varphi
_{n}\left( \xi ^{\prime }\right) t^{n}  \eqnum{2.25}
\end{eqnarray}
where

\begin{equation}
t=\exp \frac{\overline{b}}{m}\rho ,\quad \xi =\frac{p_{1}}{\sqrt{\overline{b}%
}}+\frac{x_{2\sqrt{\overline{b}}}}{2},\quad \xi ^{\prime }=\frac{p_{1}}{%
\sqrt{\overline{b}}}-\frac{x_{2\sqrt{\overline{b}}}}{2}  \eqnum{2.26}
\end{equation}

In the Landau gauge, the $\Pi _{\mu \nu }$ Euclidean components: $\Pi _{00}$%
, $\Pi _{02}$ and $\Pi _{22}$ are given by\cite{11},

\begin{mathletters}
\begin{equation}
\Pi _{00}\left( k,\mu ,\overline{b}\right) =-\frac{1}{\beta }%
\sum\limits_{p_{0}}\frac{d{\bf p}}{\left( 2\pi \right) ^{2}}G\left( p\right)
G\left( p-k\right) ,  \eqnum{2.27}
\end{equation}

\begin{equation}
\Pi _{0j}\left( k,\mu ,\overline{b}\right) =\frac{i}{2m\beta }%
\sum\limits_{p_{0}}\frac{d{\bf p}}{\left( 2\pi \right) ^{2}}\left\{ G\left(
p\right) \cdot D_{j}^{-}G\left( p-k\right) +D_{j}^{+}G\left( p\right) \cdot
G\left( p-k\right) \right\} ,  \eqnum{2.28}
\end{equation}

\end{mathletters}
\begin{eqnarray}
\Pi _{jk}\left( k,\mu ,\overline{b}\right) &=&\frac{1}{4m^{2}\beta }%
\sum\limits_{p_{0}}\frac{d{\bf p}}{\left( 2\pi \right) ^{2}}\left\{
D_{k}^{-}G\left( p\right) \cdot D_{j}^{-}G\left( p-k\right)
+D_{j}^{+}G\left( p\right) \cdot D_{k}^{+}G\left( p-k\right) \right. 
\nonumber \\
&&\left. +D_{j}^{+}D_{k}^{-}G\left( p\right) \cdot G\left( p-k\right)
+G\left( p\right) \cdot D_{j}^{-}D_{k}^{+}G\left( p-k\right) \right\} 
\nonumber \\
&&-\frac{1}{2m}\Pi _{4},  \eqnum{2.29}
\end{eqnarray}
where the notation

\begin{eqnarray}
D_{j}^{\pm }G\left( p\right) &=&\left[ ip_{j}\mp \frac{\overline{b}}{2}%
\varepsilon ^{jk}\partial _{p_{k}}\right] G\left( p\right) ,  \nonumber \\
D_{j}^{\pm }G\left( p-k\right) &=&\left[ i\left( p_{j}-k_{j}\right) \mp 
\frac{\overline{b}}{2}\varepsilon ^{jk}\partial _{p_{k}}\right] G\left(
p-k\right) ,  \eqnum{2.30}
\end{eqnarray}
has been used. The independent coefficients: ${\it \Pi }_{{\it 0}}$, ${\it %
\Pi }_{{\it 0}}\,^{\prime }$, ${\it \Pi }_{{\it 1}}$ and ${\it \Pi }_{\,{\it %
2}}$, found from (2.22)-(2.24), are functions of $k^{2}$, $\beta $ and $%
\overline{b}$.

\subsection{One-loop polarization operator coefficients in the
high-temperature approximation}

The polarization operator coefficients ${\it \Pi }_{{\it 0}}$, ${\it \Pi }_{%
{\it 0}}\,^{\prime }$, ${\it \Pi }_{{\it 1}}$ and ${\it \Pi }_{\,{\it 2}}$
corresponding to the static limit $\left( k_{0}=0,\text{ }k\rightarrow
0\right) $ in the frame ${\bf k}=(k,0)$ can be found from (2.25),
(2.27)-(22.9) through the following relations,

\begin{equation}
{\it \Pi }_{{\it 0}}+{\it \Pi }_{{\it 0}}\,^{\prime }\,k^{2}=-\Pi
^{00}\left( k_{0}=0,\text{ }k\rightarrow 0\right)  \eqnum{2.31}
\end{equation}

\begin{equation}
{\it \Pi }_{{\it 1}}k=i\Pi ^{02}\left( k_{0}=0,\text{ }k\rightarrow 0\right)
\eqnum{2.32}
\end{equation}

\begin{equation}
{\it \Pi }_{\,{\it 2}}k^{2}=\Pi _{22}\left( k_{0}=0,\text{ }k\rightarrow
0\right)  \eqnum{2.33}
\end{equation}

After summing in $p_{0}$ in eqs. (2.27)-(2.29), and using the relations
(2.31)-(2.33), we find that the polarization operator coefficients, in the $%
k/\sqrt{\overline{b}}\ll 1$ limit, are given by

\begin{equation}
{\it \Pi }_{{\it 0}}+{\it \Pi }_{{\it 0}}\,^{\prime }\,k^{2}=\frac{\beta 
\overline{b}}{8\pi }\sum_{n}\Theta _{n}+k^{2}\left[ \frac{2m}{\pi \overline{b%
}}\sum_{n}\Delta _{n}-\frac{\beta }{8\pi }\sum_{n}(2n+1)\Theta _{n}\right] 
\eqnum{2.34}
\end{equation}

\begin{equation}
{\it \Pi }_{{\it 1}}=\left[ \frac{1}{\pi }\sum_{n}\Delta _{n}-\frac{\beta 
\overline{b}}{16\pi m}\sum_{n}(2n+1)\Theta _{n}\right]  \eqnum{2.35}
\end{equation}

\begin{equation}
{\it \Pi }_{\,{\it 2}}=\left[ \frac{1}{\pi m}\sum_{n}(2n+1)\Delta _{n}-\frac{%
\beta \overline{b}}{32\pi m^{2}}\sum_{n}(2n+1)^{2}\Theta _{n}\right] 
\eqnum{2.36}
\end{equation}
where

\begin{equation}
\Theta _{n}=%
\mathop{\rm sech}%
\,^{2}\frac{\beta (\frac{\epsilon _{n}}{2}-\mu )}{2},\qquad \Delta
_{n}=(e^{\beta (\frac{\epsilon _{n}}{2}-\mu )}+1)^{-1}  \eqnum{2.37}
\end{equation}

To find the high-temperature leading contributions of the one-loop
polarization operator coefficients, the summations in (2.34)-(2.36) are
carried out using a high-temperature asymptotic expansion in the
Euler-MacLauring sum formula. Their high-temperature leading contributions
are\cite{8},\cite{14}

\begin{equation}
{\it \Pi }_{{\it 0}}=\frac{m}{2\pi }\left[ \tanh \frac{\beta \mu }{2}%
+1\right] ,\qquad {\it \Pi }_{{\it 0}}\,^{\prime }=-\frac{\beta }{48\pi }%
\mathop{\rm sech}%
\!^{2}\!\,\left( \frac{\beta \mu }{2}\right) ,\qquad {\it \Pi }_{{\it 1}}=%
\frac{\overline{b}}{m}{\it \Pi }_{{\it 0}}\,^{\prime },\qquad {\it \Pi }_{\,%
{\it 2}}=\frac{1}{12m^{2}}{\it \Pi }_{{\it 0}}  \eqnum{2.38}
\end{equation}
In these expressions $m=2m_{e}$ ($m_{e}$ is the electron mass).

\section{Magnetic response in the high-temperature approximation}

In the high temperature region, for temperatures above the energy gap $%
\left( T\gg \omega _{c}\right) $, the electrons will be energized enough to
reach the empty Landau level bands, as we mentioned above. Consequently, the
electron confinement into a completely filled band is lost, being possible
for the electrons to change their initial states. Hence, it is natural to
expect, from a heuristic point of view, that in this high temperature phase
the system cannot behave as a superconductor.

If the system is not a superconductor at $T\gg \omega _{c}$, then it has to
allow the penetration of an externally applied constant magnetic field. In
other words, no Meissner effect can take place. In this section we will show
that this is indeed the case.

\subsection{Many-Particle System Linear Equations}

To obtain the extremum equations corresponding to the anyon many-particle
system we have to consider the variational problem derived from the
effective action (2.18). This formulation is known in the literature as the
self-consistent field approximation\cite{11}.

In component form, the field equations in the presence of the induced
currents and the CS fields are

\begin{equation}
{\bf \nabla }\cdot {\bf E}=eJ_{0}  \eqnum{3.1}
\end{equation}

\begin{equation}
-\partial _{0}E_{k}+\varepsilon ^{kl}\partial _{l}B=eJ^{k}  \eqnum{3.2}
\end{equation}

\begin{equation}
\frac{eN}{2\pi }b={\bf \nabla }\cdot {\bf E}  \eqnum{3.3}
\end{equation}

\begin{equation}
\frac{eN}{2\pi }f_{0k}=\varepsilon ^{kl}\partial _{0}E_{l}+\partial _{k}B 
\eqnum{3.4}
\end{equation}
where $f_{\mu \nu }$ is the CS gauge field strength tensor, defined as $%
f_{\mu \nu }=\partial _{\mu }a_{\nu }-\partial _{\nu }a_{\mu }$, and $%
J_{ind}^{\mu }$ is the current density induced by the many-particle system.

In solving eqs. (3.1)-(3.4) we confine our analysis to gauge field
configurations which are static and uniform in the y-direction. Within this
restriction we take a gauge in which $A_{1}=a_{1}=0$. Then, the different
current density components are

\begin{equation}
J_{ind}^{0}\left( x\right) ={\it \Pi }_{{\it 0}}\left[ a_{0}\left( x\right)
+eA_{0}\left( x\right) \right] +{\it \Pi }_{{\it 0}}\,^{\prime }\partial
_{x}\left( {\cal E}+eE\right) +{\it \Pi }_{{\it 1}}\left( b+eB\right) 
\eqnum{3.5}
\end{equation}

\begin{equation}
J_{ind}^{1}\left( x\right) =0,\qquad J_{ind}^{2}\left( x\right) ={\it \Pi }_{%
{\it 1}}\left( {\cal E}+eE\right) +{\it \Pi }_{\,{\it 2}}\partial _{x}\left(
b+eB\right)  \eqnum{3.6}
\end{equation}
In the above expressions we used the following notation: ${\cal E}=f_{01}$, $%
E=F_{01}$, $b=f_{12}$ and $B=F_{12}$. Eqs. (3.5)-(3.6) play the role in the
anyon fluid of the London equations in BCS superconductivity. When the
induced currents (3.5)-(3.6) are substituted in eqs. (3.1)-(3.2) we find,
after some manipulation, a set of independent differential equations
depending on the fields $B$ and $E$, along with the zero components of the
gauge potentials, $A_{0}$ and $a_{0}$,

\begin{equation}
\omega \partial _{x}^{2}B+\alpha B=\gamma \left[ \partial _{x}E-\sigma
A_{0}\right] +\tau \,a_{0},  \eqnum{3.7}
\end{equation}

\begin{equation}
\partial _{x}B=\kappa \partial _{x}^{2}E+\eta E,  \eqnum{3.8}
\end{equation}
In obtaining these equations we have used the eqs. (3.3)-(3.4) to eliminate
the CS magnetic and electric fields in terms of the Maxwell fields

\begin{equation}
b=-\chi \partial _{x}E  \eqnum{3.9}
\end{equation}

\begin{equation}
{\cal E}=-\chi \partial _{x}B  \eqnum{3.10}
\end{equation}

The coefficients appearing in these differential equations depend on the
components of the polarization operators through the relations

\[
\omega =\frac{2\pi }{N}{\it \Pi }_{{\it 0}}\,^{\prime },\quad \alpha =-e^{2}%
{\it \Pi }_{{\it 1}},\quad \tau =e{\it \Pi }_{{\it 0}},\quad \chi =\frac{%
2\pi }{eN},\quad \sigma =-\frac{e^{2}}{\gamma }{\it \Pi }_{{\it 0}},\quad
\eta =-\frac{e^{2}}{\delta }{\it \Pi }_{{\it 1}}, 
\]

\begin{equation}
\gamma =1+e^{2}{\it \Pi }_{{\it 0}}\,^{\prime }-\frac{2\pi }{N}{\it \Pi }_{%
{\it 1}},\quad \delta =1+e^{2}{\it \Pi }_{\,{\it 2}}-\frac{2\pi }{N}{\it \Pi 
}_{{\it 1}},\quad \kappa =\frac{2\pi }{N\delta }{\it \Pi }_{\,{\it 2}}. 
\eqnum{3.11}
\end{equation}

\subsection{Field Solutions at High Temperature}

Deriving eq. (3.7) with respect to $x$ and using eqs.(3.8) and (3.10), we
obtain a higher order differential equation that involves only the electric
field,

\begin{equation}
a\partial _{x}^{4}E+d\partial _{x}^{2}E+cE=0,  \eqnum{3.12}
\end{equation}
In this equation, $a=\omega \kappa $, $d=\omega \eta +\alpha \kappa -\gamma
-\tau \kappa \chi $, and $c=\alpha \eta -\sigma \gamma -\tau \eta \chi $.

Solving (3.12) we find

\begin{equation}
E\left( x\right) =C_{1}e^{-x\xi _{1}}+C_{2}e^{x\xi _{1}}+C_{3}e^{-x\xi
_{2}}+C_{4}e^{x\xi _{2}},  \eqnum{3.13}
\end{equation}
where

\begin{equation}
\xi _{1,2}=\left[ -d\pm \sqrt{d^{2}-4ac}\right] ^{\frac{1}{2}}/\sqrt{2a} 
\eqnum{3.14}
\end{equation}

The solutions for $B$, $a_{0}$ and $A_{0}$, can be obtained using eqs.
(3.8), (3.9), (3.13) and the definition of $E$ in terms of $A_{0\text{,}}$

\begin{equation}
B\left( x\right) =-\gamma _{1}\left( C_{1}e^{-x\xi _{1}}-C_{2}e^{x\xi
_{1}}\right) -\gamma _{2}\left( C_{3}e^{-x\xi _{2}}-C_{4}e^{x\xi
_{2}}\right) +C_{5}  \eqnum{3.15}
\end{equation}

\begin{equation}
a_{0}\left( x\right) =\chi \gamma _{1}\left( C_{2}e^{x\xi
_{1}}-C_{1}e^{-x\xi _{1}}\right) +\chi \gamma _{2}\left( C_{4}e^{x\xi
_{2}}-C_{3}e^{-x\xi _{2}}\right) +C_{6}  \eqnum{3.16}
\end{equation}

\begin{equation}
A_{0}\left( x\right) =\frac{1}{\xi _{1}}\left( C_{1}e^{-x\xi
_{1}}-C_{2}e^{x\xi _{1}}\right) +\frac{1}{\xi _{2}}\left( C_{3}e^{-x\xi
_{2}}-C_{4}e^{x\xi _{2}}\right) +C_{7}  \eqnum{3.17}
\end{equation}
In the above formulas we introduced the notation $\gamma _{1}=\left( \xi
_{1}^{2}\kappa +\eta \right) /\xi _{1}$, $\gamma _{2}=\left( \xi
_{2}^{2}\kappa +\eta \right) /\xi _{2}$.

As can be seen from the magnetic field solution (3.15), the real character
of the inverse length scales (3.14) is crucial for the realization of the
Meissner effect. At temperatures much lower than the energy gap $\left( T\ll
\omega _{c}\right) $ this is indeed the case, as it was shown in our
previous works (see ref. \cite{Our}).

In the high temperature region $\left( T\gg \omega _{c}\right) $ the
polarization operator coefficients are given by eqs. (2.38). Using this
approximation, and assuming that $N=2$, together with the assumption $%
n_{e}\ll m^{2}$ (this approximation is in agreement with the typical values
found in high-T$_{C}$ superconductivity), we can calculate the coefficients $%
a$, $c$ and $d$ that define the behavior of the inverse length scales,

\begin{equation}
a\simeq \pi ^{2}{\it \Pi }_{{\it 0}}{}^{\prime }{\it \Pi }_{\,{\it 2}} 
\eqnum{3.18}
\end{equation}

\begin{equation}
c\simeq e^{2}{\it \Pi }_{{\it 0}}{}  \eqnum{3.19}
\end{equation}

\begin{equation}
d\simeq -1  \eqnum{3.20}
\end{equation}

Substituting with (3.18)-(3-20) in eq. (3.14), we obtain for the inverse
length scales in the high-temperature limit

\begin{equation}
\xi _{1}\simeq e\sqrt{{\it \Pi }_{{\it 0}}}=e\sqrt{m/2\pi }\left( \tanh 
\frac{\beta \mu }{2}+1\right) ^{\frac{1}{2}}  \eqnum{3.21}
\end{equation}

\begin{equation}
\xi _{2}\simeq \frac{1}{\pi }\left( {\it \Pi }_{\,{\it 2}}{\it \Pi }_{{\it 0}%
}\,^{\prime }\right) ^{-1/2}=i\left[ 24\sqrt{\frac{2m}{\beta }}\cosh \frac{%
\beta \mu }{2}\left( \tanh \frac{\beta \mu }{2}+1\right) ^{-\frac{1}{2}%
}\right]  \eqnum{3.22}
\end{equation}

The imaginary value of the inverse length $\xi _{2}$ is due to the fact that
at $T\gg \omega _{c}$, ${\it \Pi }_{\,{\it 2}}>0$ and ${\it \Pi }_{{\it 0}%
}\,^{\prime }<0$ (see eq. (2.38)). An imaginary $\xi _{2}$ implies that the
term $\gamma _{2}\left( C_{3}e^{-x\xi _{2}}-C_{4}e^{x\xi _{2}}\right) $, in
the magnetic field solution (3.15), does not have a damping behavior, but an
oscillating one. On the other hand, the presence of the constant coefficient 
$C_{5}$ in the magnetic field solution (3.15) means that there exists a
magnetic long-range mode. In Sec. 4 we will see how this long-range mode is
associated to the existence of a mode with zero magnetic mass, ${\cal M}%
_{1}=0$.

Nevertheless, to completely determine the characteristics of the magnetic
response in this case, it is needed to find the values of the $C^{\prime }s$
unknown coefficients which are in agreement with the problem boundary
conditions and the minimization of the system free-energy density.

\subsection{ Boundary Conditions and Stability Condition for the
Semi-infinite Sample}

In order to determine the unknown coefficients, we need to use the boundary
conditions. Henceforth we consider that the anyon fluid is confined to a
half plane $-\infty <y<\infty $ with boundary at $x=0$. The external
magnetic field is applied from the vacuum ($-\infty <x<0$). The boundary
conditions for the magnetic field are $B\left( x=0\right) =\overline{B}$ ($%
\overline{B}$ constant), and $B\left( x\rightarrow \infty \right) $ finite.
Since in eqs. (3.7)-(3.8) the magnetic field is tangled to the electric
field $E$, the boundary values of $E$ have to be taken into account in
determining the unknown coefficients. Because no external electric field is
applied, the boundary conditions for this field are, $E\left( x=0\right) =0$%
, $E\left( x\rightarrow \infty \right) $ finite. After using these
conditions it is found that,

\begin{equation}
C_{2}=0,\qquad -C_{1}=C_{3}+C_{4},\qquad C_{1}=\frac{C_{5}+\left(
C_{3}-C_{4}\right) \gamma _{2}-\overline{B}}{\gamma _{1}}  \eqnum{3.23}
\end{equation}

Introducing the new notation $\xi _{2}=i\overline{\xi }_{2}$ in eqs. (3.13),
(3.15), and using the relations (3.23), we have

\begin{equation}
E\left( x\right) =C_{1}e^{-x\xi _{1}}-C_{1}\cos x\overline{\xi }_{2}+A\sin x%
\overline{\xi }_{2}  \eqnum{3.24}
\end{equation}

\begin{equation}
B\left( x\right) =-\gamma _{1}C_{1}e^{-x\xi _{1}}+\overline{\gamma }%
_{2}\left( A\cos x\overline{\xi }_{2}+C_{1}\sin x\overline{\xi }_{2}\right)
+C_{5}  \eqnum{3.25}
\end{equation}
where

\begin{equation}
A=i\left( C_{4}-C_{3}\right) ,\qquad \overline{\gamma }_{2}=\left( \overline{%
\xi }_{2}{}^{2}\kappa -\eta \right) /\overline{\xi }_{2}  \eqnum{3.26}
\end{equation}

From the boundary conditions (3.23) it is clear that the coefficient $C_{1}$
can be determined once the asymptotic condition for the magnetic field (i.e.
the value of the coefficient $C_{5}$) and the coefficient $A$ are found.

On the other hand, the coefficients $C_{6}$ and $C_{7}$, associated with the
asymptotic configurations of the potentials $A_{0}$ and $a_{0}$
respectively, are related to $C_{5}$. It is a consequence of the fact that,
in obtaining eq. (3.12), we took the derivative of eq. (3.7). Therefore, the
solution of eq. (3.12) belongs to a wider class than the one corresponding
to eqs. (3.7)-(3.10). To exclude redundant solutions we must require that
they satisfy eq. (3.7) as a supplementary condition. Therefore, substituting
the solutions (3.13), (3.15)-(3.17) into eq. (3.7), we obtain the relation

\begin{equation}
e{\it \Pi }_{{\it 1}}C_{5}=-{\it \Pi }_{{\it 0}}\left( C_{6}+eC_{7}\right) 
\eqnum{3.27}
\end{equation}

Eq. (3.27) establishes a connection between the linear combination of the
coefficients of the long-range modes of the zero components of the gauge
potentials, $(C_{6}+eC_{7})$, and the coefficient of the long-range mode of
the magnetic field, $C_{5}$. Note that if the induced CS coefficient ${\it %
\Pi }_{{\it 1}}$, or the Debye-screening coefficient ${\it \Pi }_{{\it 0}}$
were zero, there would be no link between $C_{5}$ and $(C_{6}+eC_{7})$. This
relation between the long-range modes of $B$, $A_{0}$ and $a_{0}$ can be
interpreted as a sort of Aharonov-Bohm effect, which occurs in this system
at finite temperature\cite{Our}. At $T=0$, we have ${\it \Pi }_{{\it 0}}=0$,
and this effect disappears.

After using the boundary conditions (3.23), it follows that they are not
sufficient to determine the coefficients $C_{5}$ and $A$. We need another
physical condition from where $C_{5}$ and $A$ can be found. Since,
obviously, any meaningful solution have to be stable, the natural additional
condition to be considered is the stability equation derived from the system
free energy. With this goal we start from the free energy of the half-plane
sample

\[
{\cal F}=\frac{1}{2}\int\limits_{-L^{\prime }/2}^{L^{\prime
}/2}dy\int\limits_{0}^{L}dx\left\{ \left( E^{2}+B^{2}\right) +\frac{N}{\pi }%
a_{0}b-{\it \Pi }_{{\it 0}}\left( eA_{0}+a_{0}\right) ^{2}\right. 
\]

\begin{equation}
\left. -{\it \Pi }_{{\it 0}}\,^{\prime }\left( eE+{\cal E}\right) ^{2}-2{\it %
\Pi }_{{\it 1}}\left( eA_{0}+a_{0}\right) \left( eB+b\right) +{\it \Pi }_{\,%
{\it 2}}\left( eB+b\right) ^{2}\right\}  \eqnum{3.28}
\end{equation}
where $L$ and $L^{\prime }$ determine the two sample's lengths.

In (3.28) we have to substitute the field solutions (3.16), (3.17), (3.24)
and (3.25) together with the solutions for the CS fields (that can be found
substituting (3.24) and (3.25) in eqs. (3.9) and (3.10) respectively)

\begin{equation}
b\left( x\right) =\chi \xi _{1}C_{1}e^{-x\xi _{1}}-\chi \overline{\xi }%
_{2}\left( A\cos x\overline{\xi }_{2}+C_{1}\sin x\overline{\xi }_{2}\right) 
\eqnum{3.29}
\end{equation}

\begin{equation}
{\cal E}\left( x\right) =-\chi \xi _{1}\gamma _{1}C_{1}e^{-x\xi _{1}}+\chi 
\overline{\xi }_{2}\overline{\gamma }_{2}\left( A\sin x\overline{\xi }%
_{2}-C_{1}\cos x\overline{\xi }_{2}\right) .  \eqnum{3.30}
\end{equation}

Then, after using the boundary conditions (3.23) and the constraint equation
(3.27); it is found that the leading contribution to the free-energy density 
${\it f}=\frac{{\cal F}}{{\cal A}}$ ,\ (${\cal A}=LL^{\prime }$ being the
sample area) in the sample's length limit $(L\rightarrow \infty $, $%
L^{\prime }\rightarrow \infty )$ is given as a function of $A$ and $C_{1}$ by

\begin{equation}
f=\frac{1}{2}\left[
X_{1}A^{2}+X_{2}C_{1}^{2}+X_{3}AC_{1}+X_{4}A+X_{5}C_{1}+X_{6}\right] 
\eqnum{3.31}
\end{equation}
The coefficients $X_{i}$ are expressed in terms of the polarization operator
coefficients as

\[
X_{1}=g\overline{\gamma }_{2}^{2}+{\cal G},\qquad X_{2}=g\gamma _{1}^{2}+%
{\cal G},\qquad X_{3}=-2g\gamma _{1}\overline{\gamma }_{2},\qquad X_{4}=-2g%
\overline{B}\overline{\gamma }_{2}, 
\]

\begin{equation}
X_{5}=2g\overline{B}\gamma _{1},\qquad X_{6}=g\overline{B}^{2}  \eqnum{3.32}
\end{equation}

\begin{equation}
g=1+\frac{e^{2}{\it \Pi }_{{\it 1}}\,^{2}}{{\it \Pi }_{{\it 0}}}+e^{2}{\it %
\Pi }_{\,{\it 2}}  \eqnum{3.33}
\end{equation}

\begin{eqnarray}
{\cal G} &=&\frac{1}{2}\left( 1+\overline{\gamma }_{2}^{2}-\frac{N}{\pi }%
\chi ^{2}\overline{\xi }_{2}\overline{\gamma }_{2}\right) -\left( \frac{{\it %
\Pi }_{{\it 0}}}{2\overline{\xi }_{2}^{2}}+\frac{{\it \Pi }_{{\it 0}%
}\,^{\prime }}{2}\right) \left( e+\chi \overline{\xi }_{2}\overline{\gamma }%
_{2}\right) ^{2}  \nonumber \\
&&-\frac{{\it \Pi }_{{\it 1}}}{\overline{\xi }_{2}}\left( \chi \overline{\xi 
}_{2}-e\overline{\gamma }_{2}\right) \left( e+\chi \overline{\xi }_{2}%
\overline{\gamma }_{2}\right) +\frac{{\it \Pi }_{\,{\it 2}}}{2}\left( \chi 
\overline{\xi }_{2}-e\overline{\gamma }_{2}\right) ^{2}  \eqnum{3.34}
\end{eqnarray}

The values of $A$ and $C_{1}$ are found by minimizing the corresponding
free-energy density

\begin{equation}
\frac{\delta {\it f}}{\delta A}=\frac{1}{2}\left(
2X_{1}A+X_{3}C_{1}+X_{4}\right) =0  \eqnum{3.35}
\end{equation}

\begin{equation}
\frac{\delta {\it f}}{\delta C_{1}}=\frac{1}{2}\left(
2X_{2}C_{1}+X_{3}A+X_{5}\right) =0,  \eqnum{3.36}
\end{equation}
to be

\begin{equation}
A=\frac{\overline{\gamma }_{2}}{\gamma _{1}^{2}+\overline{\gamma }_{2}^{2}}%
\overline{B}  \eqnum{3.37}
\end{equation}

\begin{equation}
C_{1}=-\frac{g\gamma _{1}^{3}}{\left( g\gamma _{1}^{2}+{\cal G}\right)
\left( \gamma _{1}^{2}+\overline{\gamma }_{2}^{2}\right) }\overline{B} 
\eqnum{3.38}
\end{equation}

Taking into account the boundary conditions (3.23) we have that the
long-range mode of the magnetic field $C_{5}$ is given by

\begin{equation}
C_{5}=\gamma _{1}C_{1}-\overline{\gamma }_{2}A+\overline{B}=\frac{\gamma
_{1}^{2}{\cal G}}{\left( g\gamma _{1}^{2}+{\cal G}\right) \left( \gamma
_{1}^{2}+\overline{\gamma }_{2}^{2}\right) }\overline{B}  \eqnum{3.39}
\end{equation}

From (3.39) we see that for $T\gg \omega _{c}$ the electromagnetic field
long-range mode propagates into the sample, producing an homogeneous
magnetic penetration. This result is different from the one obtained in the
low-temperature limit ($T\ll \omega _{c}$) \cite{Our}. In that limit it was
found that $C_{5}=0$, which implies that the long-range mode cannot
propagate within the sample when a uniform and constant magnetic field is
perpendicularly applied at the sample's boundaries.

\subsection{Inhomogeneous Magnetic Response}

As it has been previously established, in the high-temperature limit the
coefficients $A$, $C_{1}$ and $C_{5}$ are all different from zero, i.e., in
the ($T\gg \omega _{c}$)-phase the magnetic response of the charged anyon
fluid has an exponential decaying component, as well as, both homogeneous
and inhomogeneous penetrations.. To complete our study of the magnetic
response at high temperature we still need to estimate the corresponding
values of each component for the range of parameters and temperatures here
considered.

At the densities under consideration, $n_{e}\ll m^{2}$, the estimated values
of the coefficients $A$, $C_{1}$ and $C_{5}$ in the high-temperature
approximation $\left( T\gg \omega _{c}\right) $ are

\begin{equation}
A\approx 10^{3}\overline{B},\qquad C_{1}\approx -10^{-11}\overline{B},\qquad
C_{5}\approx 10^{-4}\overline{B}  \eqnum{3.40}
\end{equation}
In this approximation the leading contributions to the electric and the
magnetic fields, (3.24) and (3.25), are then given respectively by

\begin{equation}
E\left( x\right) =E_{0}\left( T\right) \sin \left( \frac{2\pi }{\lambda }%
x\right)  \eqnum{3.41}
\end{equation}

\begin{equation}
B\left( x\right) =\overline{B}\cos \left( \frac{2\pi }{\lambda }x\right) 
\eqnum{3.42}
\end{equation}
where

\begin{equation}
E_{0}\left( T\right) =\frac{12\sqrt{2}m}{\overline{\xi }_{2}}\left( \tanh 
\frac{\beta \mu }{2}+1\right) ^{-1}\overline{B}  \eqnum{3.43}
\end{equation}

\begin{equation}
\lambda =\frac{2\pi }{\overline{\xi }_{2}}  \eqnum{3.44}
\end{equation}

From eq. (3.42), one sees that the magnetic response exponentially decaying
component of the (the one associated with the coefficient $\gamma _{1}C_{1}$
in the general solution (3.25)), as well as the uniform one (3.39), are
negligible if compared with the inhomogeneous component associated with the
coefficient $A$.

Hence, at $T\gg \omega _{c}$ the applied magnetic field penetrates the
charged anyon fluid with a magnitude that changes sinusoidally with $x$ and
has an amplitude $\overline{B}$. Moreover, the inhomogeneous magnetic field
penetration (3.42) is characterized by a wavelength $\lambda $, which is
proportional to the inverse of the length scale magnitude $\overline{\xi }%
_{2}$ (eq. (3.44)).

At $T\gtrsim \omega _{c}$, using that $\mu \simeq 
{\displaystyle {\pi n_{e} \over m}}%
$ \cite{8}, one can estimate from (3.44) and (3.22) that $\lambda \simeq 0.4$
$A^{o}$. On the other hand, taking into account that $\overline{\xi }_{2}$
increases with the temperature (see eq. (3.22)), we have, that the
wavelength decreases with $T$. The high-temperature leading behavior for $%
\lambda $ is given by

\begin{equation}
\lambda \approx \frac{\pi }{24}\sqrt{\frac{1}{2mT}}  \eqnum{3.45}
\end{equation}

Note that when an external constant and uniform magnetic field is applied to
the charged anyon fluid in the $\left( T\gg \omega _{c}\right) $-phase, an
inhomogeneous electric field (3.24) is induced within the medium. The
amplitude of this induced electric field depends on the magnitude of the
applied magnetic field, $\overline{B}$, and the temperature. The $E$'s
inhomogeneity also increases with the temperature through $\lambda $.

From the obtained results we conclude that for temperatures larger than the
energy gap, the charged anyon fluid is in a new phase on which the
superconductivity is lost (non Meissner effect is found in this phase).

The induction of inhomogeneous electric and magnetic fields within the
charged anyon fluid at high temperature, indicates that some redistribution
of the induced charge and currents occurs at $T\gg \omega _{c}$.

To verify this, let us calculate the induced electric charge density of the
charged medium in the high-temperature approximation. Considering eq. (3.5)
in the high-temperature limit, we find that the induced electric charge
density presents an inhomogeneous spatial distribution with high-temperature
leading contribution given by

\begin{equation}
eJ_{0}\left( x\right) =24\sqrt{2}m\overline{B}\left[ \tanh \left( \frac{%
\beta \mu }{2}\right) +1\right] ^{-1}\cos \left( \frac{2\pi }{\lambda }%
x\right)  \eqnum{3.46}
\end{equation}

As discussed above, in the high-temperature regime $\lambda \sim \sqrt{1/T}$%
, so the spatial inhomogeneity of the charge density (3.46) increases with
the temperature.

In the same way, if we calculate the current density (3.6) in the
high-temperature limit we find

\begin{equation}
eJ_{2}\left( x\right) =-96\pi \sqrt{2}\sqrt{m/\beta }\overline{B}\cosh
\left( \frac{\beta \mu }{2}\right) \left[ \tanh \left( \frac{\beta \mu }{2}%
\right) +1\right] ^{-\frac{1}{2}}\sin \left( \frac{2\pi }{\lambda }x\right) 
\eqnum{3.47}
\end{equation}

Obviously, the current density (3.47) is not a supercurrent confined to the
sample's boundary.

\section{Magnetic mass and rest energy in the charged anyon fluid at $T\gg
\omega _{c}$}

We have seen that the inverse length scales, $\xi _{1}$ and $\xi _{2}$, are
basic elements in the determination of the magnetic response of the charged
anyon fluid. In this Sec. we shall go one step forward in clarifying the
physical interpretation of these parameters. We will show that the inverse
length scales (3.21),(3.22) can be identified with the magnetic masses of
the electromagnetic field within the fluid at $T\gg \omega _{c}$. A
particularly important point in this Sec. is our proof that the existence at
high temperature of an imaginary magnetic mass (corresponding to the inverse
length (3.22)) is not linked to the presence of tachyons in the theory, or
to the breaking of the linear approximation, as was suggested in ref. \cite
{8}. As shown below, the magnetic mass and the rest energy are not the same
in the MCS theory (contrary to what happens in a Klein-Gordon-like theory).

To investigate the magnetic masses and the rest energies associated with the
electromagnetic modes, we need to study the electromagnetic field dispersion
equation. With this aim we start from the effective action (2.18) taken in
the covariant gauges for the Maxwell and CS fields

\begin{equation}
\frac{1}{\alpha _{1}}\partial _{\mu }A^{\mu }=0,\qquad \frac{1}{\alpha _{2}}%
\partial _{\mu }a^{\mu }=0  \eqnum{4.1}
\end{equation}
$\alpha _{1}$ and $\alpha _{2}$ being two independent gauge parameters.

The corresponding effective Lagrangian density for the Maxwell and CS field
configurations can be represented as

\begin{equation}
{\cal L}_{eff}=-\frac{1}{2}A^{\mu }\left( -k\right) \Delta _{\mu \nu
}^{-1}\left( k\right) A^{\nu }\left( k\right) -\frac{1}{2}a^{\mu }\left(
-k\right) {\cal D}_{\mu \nu }^{-1}\left( k\right) a^{\nu }\left( k\right)
-eA^{\mu }\left( -k\right) \Pi _{\mu \nu }\left( k\right) a^{\nu }\left(
k\right)  \eqnum{4.2}
\end{equation}
where the matrices $\Delta _{\mu \nu }^{-1}$ and ${\cal D}_{\mu \nu }^{-1}$
are given by

\begin{equation}
\Delta _{\mu \nu }^{-1}\left( k\right) =k^{2}g_{\mu \nu }-\left( 1-\frac{1}{%
\alpha _{1}}\right) k_{\mu }k_{\nu }+e^{2}\Pi _{\mu \nu }\left( k\right) 
\eqnum{4.3}
\end{equation}

\begin{equation}
{\cal D}_{\mu \nu }^{-1}\left( k\right) =\frac{iN}{2\pi }\epsilon _{\mu \rho
\nu }k^{\rho }+\frac{1}{\alpha _{2}}k_{\mu }k_{\nu }+\Pi _{\mu \nu }\left(
k\right)  \eqnum{4.4}
\end{equation}
$\Pi _{\mu \nu }$ is the one-loop fermion polarization operator given from
(2.20)-(2.21) by

\begin{equation}
\Pi ^{\mu \nu }=\left( 
\begin{array}{ccc}
-{\cal A}_{1}\,k^{2} & {\cal A}_{1}k\omega & -i{\it \Pi }_{{\it 1}}k \\ 
{\cal A}_{1}k\omega & -{\cal A}_{1}\omega ^{2} & i{\it \Pi }_{{\it 1}}\omega
\\ 
i{\it \Pi }_{{\it 1}}k & -i{\it \Pi }_{{\it 1}}\omega & -{\cal A}_{1}\omega
^{2}+{\it \Pi }_{\,{\it 2}}k^{2}
\end{array}
\right)  \eqnum{4.5}
\end{equation}
In (4.5) we are considering the frame ${\bf k}=(k,0)$, and the notation $%
\omega =k_{0}$ has been used.

The effective theory for the electromagnetic field in the charged anyon
fluid, is found integrating the CS fields in the partition function (2.3)
with Lagrangian density (4.2). The new effective Lagrangian density so
obtained is

\begin{equation}
{\cal L}_{eff}^{\prime }=-\frac{1}{2}A^{\mu }\left( -k_{\rho }\right) \left[
\Delta _{\mu \nu }^{-1}\left( k_{\rho }\right) -N_{\mu \nu }\left( k_{\rho
}\right) \right] A^{\nu }\left( k_{\rho }\right)  \eqnum{4.6}
\end{equation}
where the matrix $N_{\mu \nu }\left( k_{\rho }\right) $ is defined by

\begin{equation}
N^{\mu \nu }\left( k_{\rho }\right) =e^{2}\Pi ^{\mu \lambda }\left( k_{\rho
}\right) {\cal D}_{\lambda \rho }\left( k_{\rho }\right) \Pi ^{\rho \nu
}\left( k_{\rho }\right)  \eqnum{4.7}
\end{equation}

The dispersion equation for the Maxwell field is then given by

\begin{equation}
\det \left[ \Delta _{\mu \nu }^{-1}\left( \omega ,\text{ }k\right) -N_{\mu
\nu }\left( \omega ,\text{ }k\right) \right] =0  \eqnum{4.8}
\end{equation}

It is noteworthy to discuss now some of the characteristics of the
dispersion equation in QFT. In general, the dispersion equation can be
reduced to an equation of the form

\begin{equation}
F_{1}\left( k,\text{ }\omega \right) +F_{2}\left( k,\text{ }\omega \right) -%
{\cal M}^{2}=0  \eqnum{4.9}
\end{equation}
where $F_{1}\left( k,\text{ }\omega \right) $ is a homogeneous function in $%
\omega $ and $F_{2}\left( k,\text{ }\omega \right) $ is a homogeneous
function in $k$. The squared magnetic mass is then defined as the solution
for $-k^{2}$ found from the dispersion equation (4.9) evaluated at $\omega
=0 $,

\begin{equation}
F_{2}\left( k,\omega =0\right) -{\cal M}^{2}=0  \eqnum{4.10}
\end{equation}
while the system rest energy (from where the existence of tachyons can be
determined) has to be found as the $\omega $ solution of (4.9) when $k=0$,

\begin{equation}
F_{1}\left( \omega ,k=0\right) -{\cal M}^{2}=0  \eqnum{4.11}
\end{equation}
In a Klein-Gordon-like theory, $F_{1}\left( k,\text{ }\omega \right) =\omega
^{2}$ and $F_{2}\left( k,\text{ }\omega \right) =-k^{2}$, then the magnetic
mass is just equal to the rest energy and given by ${\cal M}$, but in
general this does not have to necessarily be the case. For example, in
relativistic systems of charged fermions at finite density (i.e. in the
presence of a chemical potential $\mu $) the dispersion equation has a
linear term in $\omega $\cite{Fradkyn},

\begin{equation}
\omega ^{2}+\mu \omega -\left( k^{2}+{\cal M}^{2}\right) =0  \eqnum{4.12}
\end{equation}
In that case the solution for $-k^{2}$ with $\omega =0$ is equal to ${\cal M}
$, while, as it is known, the rest energy for particles and antiparticles is
given by

\begin{equation}
\omega =-\frac{\mu }{2}\pm \sqrt{\left( \frac{\mu }{2}\right) ^{2}+{\cal M}%
^{2}}  \eqnum{4.13}
\end{equation}

Furthermore, in the presence of a background magnetic field, the distinction
between mass and rest energy becomes essential. For instance, in the context
of string theory in a background magnetic field, it has been shown\cite
{String} that the mass (defined in agreement with Wigner's definition) of
higher spin ($s\geq 1$) charged boson particles does not coincide with their
rest energy. This is due to the modification of the algebra of the global
symmetries by the background field.

From the above discussion, it is also clear that in calculating the magnetic
mass (eq. (4.10)) and the rest energy (eq. (4.11)) at finite temperature, we
have to take the polarization operators coefficients in the static limit ($%
\omega =0$, $k\sim 0$) and in the plasmon limit ($k=0$, $\omega \sim 0$)
respectively. Now, because of the lack of analyticity of the Green's
function about $k_{\mu }=0$ at $T\neq 0$, it is known that in QFT these
limits do not commute\cite{Fradkyn}, \cite{Linde}. In anyon theory at $T\neq
0$ one faces a similar situation, as it was shown in ref. \cite{14}. Then,
in each case we have to consider the polarization operator coefficients
evaluated in the corresponding limit.

In anyon theory the CS interaction gives rise to a dispersion equation with
a structure more complicated than that corresponding to a Klein-Gordon-like
theory. As shown below, in this case the magnetic masses and the rest
energies of the electromagnetic modes are different.

\subsection{Electromagnetic field magnetic masses at high temperatures}

To find the electromagnetic field magnetic masses we must solve the
dispersion equation (4.8) at $\omega =0$. Since we are interested in the
magnetic masses at temperatures higher than the energy gap $\left( T\gg
\omega _{c}\right) $, we should consider in solving (4.8) the polarization
operator coefficients in the high-temperature approximation (2.38).

Let us determine first the expression of the matrix $N^{\mu \nu }\left(
k\right) $ (eq. (4.7)) at $\omega =0$. In this case the matrix ${\cal D}%
_{\lambda \rho }\left( k\right) $ appearing in eq. (4.7) takes the form

\begin{equation}
{\cal D}_{\lambda \rho }\left( k\right) =\frac{k^{2}}{\alpha _{2}{\it D}}%
\left[ 
\begin{array}{ccc}
{\it \Pi }_{\,{\it 2}}k^{2} & 0 & iHk \\ 
0 & -\left( {\it \Pi }_{\,{\it 2}}{\cal A}_{1}k^{2}+H^{2}\right) \alpha _{2}
& 0 \\ 
-iHk & 0 & -{\cal A}_{1}k^{2}
\end{array}
\right]  \eqnum{4.14}
\end{equation}
In writing (4.14) we have used the following notation,

\begin{equation}
{\it D}=\det {\cal D}_{\mu \nu }^{-1}\left( k\right) =-k^{4}\left( {\it \Pi }%
_{\,{\it 2}}{\cal A}_{1}k^{2}+H^{2}\right) /\alpha _{2}  \eqnum{4.15}
\end{equation}

\begin{equation}
H=-\frac{1}{\pi }+{\it \Pi }_{{\it 1}}  \eqnum{4.16}
\end{equation}
Thus, after taking the matrix products indicated in (4.7) with the
polarization operator (4.5) evaluated at $\omega =0$, we find,

\begin{equation}
N^{\mu \nu }\left( k\right) =\frac{e^{2}}{\alpha _{2}{\it D}}\left[ 
\begin{array}{ccc}
{\cal A}_{1}Bk^{2} & 0 & ikC \\ 
0 & 0 & 0 \\ 
-ikC & 0 & -{\it \Pi }_{\,{\it 2}}Bk^{2}
\end{array}
\right]  \eqnum{4.17}
\end{equation}
with

\begin{equation}
B=\left( {\it \Pi }_{\,{\it 2}}{\cal A}_{1}k^{2}+2H{\it \Pi }_{{\it 1}}-{\it %
\Pi }_{{\it 1}}^{\,\ \;2}\right) k^{4},\qquad C=\left[ {\cal A}%
_{1}k^{2}\left( 2{\it \Pi }_{{\it 1}}{\it \Pi }_{\,{\it 2}}-H{\it \Pi }_{\,%
{\it 2}}\right) +H{\it \Pi }_{{\it 1}}^{\,\ \;2}\right] k^{4}  \eqnum{4.18}
\end{equation}

Using (4.3) and (4.17) in the dispersion equation (4.8) we obtain

\[
\det \left[ \Delta _{\mu \nu }^{-1}\left( \omega =0,\;k\right) -N_{\mu \nu
}\left( \omega =0,\;k\right) \right] = 
\]

\begin{equation}
=\frac{-k^{4}}{\alpha _{1}}\left[ \left( k^{2}+e^{2}{\cal A}%
_{1}G_{1}k^{2}\right) \left( 1+e^{2}G_{1}{\it \Pi }_{\,{\it 2}}\right)
+e^{4}G_{2}^{2}\right] =0  \eqnum{4.19}
\end{equation}
with

\begin{equation}
G_{1}=\frac{B}{\alpha _{2}{\it D}}+1\text{,\qquad }  \eqnum{4.20}
\end{equation}

\begin{equation}
G_{2}=\frac{C}{\alpha _{2}{\it D}}+{\it \Pi }_{{\it 1}}  \eqnum{4.21}
\end{equation}

For $N=2$, taking into account that in natural units $e^{2}\sim
10^{5}cm^{-1} $, and considering the characteristic values $n_{e}=2\times
10^{14}cm^{-2}$ and $m_{e}=2.6\times 10^{10}cm^{-1}$, we can estimate the
relative orders between the polarization operator coefficients (2.38) at
temperatures larger than $\omega _{c}$ as

\begin{equation}
{\it \Pi }_{{\it 0}}\approx -10^{8}e^{2}{\it \Pi }_{{\it 1}}\approx
-10^{7}e^{4}{\it \Pi }_{{\it 0}}\,^{\prime }\approx 10^{20}e^{4}{\it \Pi }%
_{\,{\it 2}}  \eqnum{4.22}
\end{equation}

Taking into account the relations (4.22), the high-temperature leading
contribution to the dispersion equation (4.19) is

\begin{equation}
\det \left[ \Delta \left( k\right) -N\left( k\right) \right] \simeq -\frac{%
\left( {\it \Pi }_{\,{\it 2}}{\it \Pi }_{{\it 0}}\,^{\prime }\right) ^{2}}{%
\alpha _{1}\left( \alpha _{2}{\it D}\right) ^{2}}k^{12}\left( k^{6}+\sigma
_{1}k^{4}+\sigma _{2}k^{2}+\sigma _{3}\right) =0  \eqnum{4.23}
\end{equation}
where

\begin{equation}
\sigma _{1}=\frac{2}{\pi ^{2}}\left( {\it \Pi }_{\,{\it 2}}{\it \Pi }_{{\it 0%
}}\,^{\prime }\right) ^{-1},\qquad \sigma _{2}=\frac{\sigma _{1}^{2}}{4}%
,\qquad \sigma _{3}=\frac{\sigma _{1}^{2}}{4}e^{2}{\it \Pi }_{{\it 0}} 
\eqnum{4.24}
\end{equation}

To find the magnetic masses of the electromagnetic field in the charged
anyon fluid we need to find the zeros in $k$ of the dispersion equation
(4.23). In doing that, we have to solve a cubic equation in $k^{2}$ (see the
polynomial into the parenthesis in eq. (4.23))

\begin{equation}
y^{3}+\sigma _{1}y^{2}+\sigma _{2}y+\sigma _{3}=0  \eqnum{4.25}
\end{equation}
In (4.25) we made the variable change $y=k^{2}$. The roots of (4.25) are
given by

\begin{equation}
y_{1}=A+B-\frac{\sigma _{1}}{3},\qquad y_{2,3}=-\frac{A+B}{2}\pm i\frac{A-B}{%
2}\sqrt{3}-\frac{\sigma _{1}}{3}  \eqnum{4.26}
\end{equation}
with

\[
A=\frac{\sigma _{1}}{6}\left( 1-\vartheta _{A}\right) ^{1/3},\qquad B=\frac{%
\sigma _{1}}{6}\left( 1-\vartheta _{B}\right) ^{1/3},\qquad \vartheta
_{A,B}=3^{3}\vartheta \mp 3i\sqrt{6\vartheta }, 
\]

\begin{equation}
\vartheta =\frac{e^{2}{\it \Pi }_{{\it 0}}}{\sigma _{1}}  \eqnum{4.27}
\end{equation}
In the high-temperature limit $\vartheta _{A,B}\ll 1$, so we can take the
expansions,

\begin{equation}
\left( 1-\vartheta _{A,B}\right) ^{1/3}\approx 1-\frac{1}{3}\vartheta _{A,B}-%
\frac{1}{9}\vartheta _{A,B}^{2}  \eqnum{4.28}
\end{equation}
Substituting with (4.27) and (4.28) in (4.26) we obtain the leading
high-temperature approximation for the roots of eq. (4.25),

\begin{equation}
y_{1}=-e^{2}{\it \Pi }_{{\it 0}},\qquad y_{2,3}=-\frac{1}{\pi ^{2}}\left( 
{\it \Pi }_{\,{\it 2}}{\it \Pi }_{{\it 0}}\,^{\prime }\right) ^{-1} 
\eqnum{4.29}
\end{equation}

The solutions of (4.23) obtained in the high-temperature approximation
represents the electromagnetic field magnetic mass spectrum

\begin{equation}
k^{2}+{\cal M}_{j}^{2}=0,\qquad j=1,2,3  \eqnum{4.30}
\end{equation}

The squared magnetic masses in the high-temperature approximation are found
from (4.29) to be

\begin{equation}
{\cal M}_{1}^{2}=0  \eqnum{4.31}
\end{equation}

\begin{equation}
{\cal M}_{2}^{2}=e^{2}{\it \Pi }_{{\it 0}}  \eqnum{4.32}
\end{equation}

\begin{equation}
{\cal M}_{3}^{2}=\frac{1}{\pi ^{2}}\left( {\it \Pi }_{\,{\it 2}}{\it \Pi }_{%
{\it 0}}\,^{\prime }\right) ^{-1}  \eqnum{4.33}
\end{equation}

We should note that the magnetic masses (4.32)-(4.33) are gauge independent.
That is, they do not depend on the gauge parameters $\alpha _{1}$ and $%
\alpha _{2}$. We can see that the equation (4.23), from where the magnetic
masses are found, is independent of the gauge parameter $\alpha _{2}$, since 
$\alpha _{2}{\it D}$ does not depend on $\alpha _{2}$; and $\alpha _{1}$
appears only as a multiplicative factor.

From (4.31)-(4.33) we have that one of the infrared modes of the
electromagnetic field in the anyon fluid is massless, ${\cal M}_{1}=0$,
while ${\cal M}_{2}^{2}>0$ and ${\cal M}_{3}^{2}<0$. The signs obtained for
the square of the masses (4.32), (4.33), are a consequence of the fact that
in the high-temperature limit, ${\it \Pi }_{{\it 0}}>0$, ${\it \Pi }_{\,{\it %
2}}>0$ and ${\it \Pi }_{{\it 0}}\,^{\prime }<0$ (eq. (2.38)).

Comparing eqs. (4.32), (4.33) with eqs. (3.21), (3.22) respectively, one can
see that

\begin{equation}
\xi _{1}={\cal M}_{2},\qquad \xi _{2}={\cal M}_{3}  \eqnum{4.34}
\end{equation}

Hence, the magnetic masses coincide with the inverse length scales, $\xi
_{1} $ and $\xi _{2}$, which determine the magnetic response of the medium.
This is precisely the physical meaning of these infrared masses (${\cal M}%
_{i}$). They cannot be interpreted, otherwise, as the rest energies of the
electromagnetic field modes, as we will see below. The zero mode (4.31) is
linked to the long-range component $C_{5}$ appearing in the magnetic
response (3.25).

\subsection{Electromagnetic field rest energies at high temperatures}

The rest energies of the electromagnetic modes are found by solving the
dispersion equation (4.8) for $\omega $ at $k=0$. We recall that the
polarization operator coefficients have to be taken now in the plasmon limit
($k=0$, $\omega \sim 0$). Using this limit, we find that ${\it \Pi }_{{\it 0}%
}\left( k=0,\omega \sim 0\right) =0$, while the rest of the coefficients
maintain the same functional behavior (2.38) obtained in the static limit 
\cite{14}. Then the polarization operator is given by (4.5) with ${\it \Pi }%
_{{\it 0}}=0$ and $k=0$.

In this case the matrix ${\cal D}_{\lambda \rho }\left( \omega \right) $
appearing in eq. (4.7) takes the form

\begin{equation}
{\cal D}^{\lambda \rho }\left( \omega \right) =\frac{1}{{\it D}}\left[ 
\begin{array}{ccc}
\frac{\alpha _{2}}{\omega ^{2}}{\it D} & 0 & 0 \\ 
0 & {\cal -}{\it \Pi }_{{\it 0}}\,^{\prime }\frac{\omega ^{4}}{\alpha _{2}}
& -iH\frac{\omega ^{3}}{\alpha _{2}} \\ 
0 & iH\frac{\omega ^{3}}{\alpha _{2}} & {\cal -}{\it \Pi }_{{\it 0}%
}\,^{\prime }\frac{\omega ^{4}}{\alpha _{2}}
\end{array}
\right]  \eqnum{4.35}
\end{equation}
In writing (4.35) the following notation was used

\begin{equation}
{\it D}=\det {\cal D}_{\mu \nu }^{-1}\left( k\right) =\frac{\omega ^{4}}{%
\alpha _{2}}\left[ \left( {\it \Pi }_{{\it 0}}\,^{\prime }\right) ^{2}\omega
^{2}-H^{2}\right]  \eqnum{4.36}
\end{equation}

Thus, after taking the matrix products indicated in (4.7) we find,

\begin{equation}
N^{\mu \nu }\left( \omega \right) =\frac{e^{2}}{\alpha _{2}{\it D}}\left[ 
\begin{array}{ccc}
0 & 0 & 0 \\ 
0 & M & N \\ 
0 & -N & M
\end{array}
\right]  \eqnum{4.37}
\end{equation}
with

\begin{equation}
M=-\omega ^{6}{\it \Pi }_{{\it 0}}\,^{\prime }\left[ \left( {\it \Pi }_{{\it %
0}}\,^{\prime }\right) ^{2}\omega ^{2}+\left( {\it \Pi }_{{\it 1}}-2H\right) 
{\it \Pi }_{{\it 1}}\right] ,  \eqnum{4.38}
\end{equation}

\begin{equation}
N=i\omega ^{5}\left[ \left( 2{\it \Pi }_{{\it 1}}-H\right) \left( {\it \Pi }%
_{{\it 0}}\,^{\prime }\right) ^{2}\omega ^{2}-H\left( {\it \Pi }_{{\it 1}%
}\right) ^{2}\right]  \eqnum{4.39}
\end{equation}

Using (4.3) and (4.37) in the dispersion equation (4.8) we obtain the
general expression for the rest-energy equation

\[
\det \left[ \Delta _{\mu \nu }^{-1}\left( \omega \right) -N_{\mu \nu }\left(
\omega \right) \right] =\frac{\omega ^{2}}{\alpha _{1}}\cdot \left\{ \left[
\left( 1+e^{2}{\it \Pi }_{{\it 0}}\,^{\prime }\right) \omega ^{2}+\frac{e^{2}%
}{\alpha _{2}{\cal D}}M\right] ^{2}+\right. 
\]

\begin{equation}
\text{ }\left. \left[ ie^{2}{\it \Pi }_{{\it 1}}\omega -\frac{e^{2}}{\alpha
_{2}{\cal D}}N\right] ^{2}\right\} =0  \eqnum{4.40}
\end{equation}
Taking into account the relations (4.22), the high-temperature leading
contribution to the rest-energy equation (4.40) is

\begin{equation}
\det \left[ \Delta ^{-1}\left( k\right) -N\left( k\right) \right] \simeq 
\frac{\left( {\it \Pi }_{{\it 0}}\,^{\prime }\right) ^{4}\omega ^{10}}{%
\alpha _{1}\left( \alpha _{2}{\it D}\right) ^{2}}\left( \omega ^{6}+\theta
_{1}\omega ^{4}+\theta _{2}\omega ^{2}+\theta _{3}\right) =0  \eqnum{4.41}
\end{equation}
where

\begin{equation}
\theta _{1}=-\frac{2H^{2}}{\left( {\it \Pi }_{{\it 0}}\,^{\prime }\right)
^{2}},\qquad \theta _{2}=\frac{1}{4}\theta _{1}^{2},\qquad \theta _{3}=-%
\frac{1}{4}\theta _{1}^{2}\left( e^{2}{\it \Pi }_{{\it 1}}\right) ^{2} 
\eqnum{4.42}
\end{equation}

The $\omega $ solutions of eq. (4.41) are

\begin{equation}
\omega _{1}^{2}=0,  \eqnum{4.43}
\end{equation}

\begin{equation}
\omega _{2}^{2}=\left( e^{2}{\it \Pi }_{{\it 1}}\right) ^{2}  \eqnum{4.44}
\end{equation}

\begin{equation}
\omega _{3}^{2}=\left( \frac{N}{2\pi {\it \Pi }_{{\it 0}}\,^{\prime }}%
\right) ^{2}  \eqnum{4.45}
\end{equation}

From (4.43)-(4.45) we have that $\omega _{2,3}^{2}>0$; therefore, at $T\gg
\omega _{c}$ there is no negative squared rest energy, which means that the
high temperature phase is stable. Hence, the existence of a negative squared
magnetic mass, ${\cal M}_{3}^{2}$, simply indicates that there is an
inhomogeneous magnetic penetration in the charged anyon fluid at $T\gg
\omega _{c}$.

Finally, we should point out that the rest energies $\omega _{2,3}$ are also
gauge independent (they do not depend on $\alpha _{1}$ and $\alpha _{2}$),
and they are determined by the explicit (proportional to $N$) and induced
(proportional to ${\it \Pi }_{{\it 1}}$) CS contributions.

\section{Concluding Remarks}

The particle energy spectrum of the anyon theory exhibits a band structure
given by different Landau levels separated by an energy gap $\omega _{c}$.
The energy gap is proportional to the background CS magnetic field $%
\overline{b}$, which is induced in the charged anyon fluid to guarantee the
electrical neutrality of the system. As it was shown in our previous works 
\cite{Our}, at temperatures lower than the energy gap ($T\ll \omega _{c}$) a
constant and uniform applied magnetic field cannot penetrate the anyon fluid
(i.e. the Meissner effect takes place in that superconducting phase).

In this paper we have proved that at $T\gg \omega _{c}$ the charged anyon
fluid does not exhibit a Meissner effect. Hence, we can conclude that the
energy gap $\omega _{c}$ defines a scale that separates two phases in the
charged anyon fluid: a superconducting phase at $T\ll \omega _{c}$, and a
non-superconducting one at $T\gg \omega _{c}$. We expect that the critical
temperature for this phase transition should be of order $\omega _{c}$.
Nevertheless, the temperature approximation (2.38) is not suitable to
perform the calculation needed to find the phase transition temperature.

We must emphasize that the scenario we have found here and in previous works
for the anyon superconductivity at finite temperature is in agreement with
the heuristic understanding of anyon superconductivity given by Wilczek in
ref. \cite{Wilczek}. There, Wilczek pointed out that the London arguments,
which start from the role of the energy gap as an essential fact in the
theory of superconductivity, seem to provide the base for anyon
superconductivity. In the charged anyon fluid, there is no the
charge-violating local order parameter which is familiar in the theories
with spontaneously broken symmetry. In this system, instead, it is the
background CS magnetic field $\overline{b}$ what determines the energy gap ($%
\omega _{c}=\overline{b}/m$) and plays the role of the order parameter in
the anyon gas \cite{Chen}. Then, it is natural to expect that the critical
temperature of the superconducting phase is associated with the anyon fluid
order parameter, i.e. with the energy gap.

As we have found in this paper, at $T\gg \omega _{c}$ the magnetic response
of the charged anyon fluid exhibits an inhomogeneous penetration
characterized by a wavelength which decreases with the temperature as $1/%
\sqrt{T}$ in its high-temperature leading order. That is, the spatial
inhomogeneity of the many-particle magnetic response will increase with the
temperature in this new phase.

The absence of the Meissner effect at $T\gg \omega _{c}$ is related to the
appearing of an imaginary magnetic mass ${\cal M}_{3}$ in this phase (eq.
(4.33)). The existence of an imaginary magnetic mass cannot be associated
with a tachyonic mode in this many-particle system with CS interactions. The
reason is that the magnetic masses and the rest energies of the
electromagnetic field modes are not the same within the charged anyon fluid.

It is important to note that in obtaining the inhomogeneous magnetic
response at high temperatures, it was crucial that the polarization operator
coefficient ${\it \Pi }_{{\it 0}}{}^{\prime }$ changed its sign from a
positive value at $T\ll \omega _{c}$ \cite{8},\cite{14},\cite{Our}, to a
negative value at $T\gg \omega _{c}$ (eq. (2.38)). That is, because ${\it %
\Pi }_{{\it 0}}{}^{\prime }$ changes its sign, while ${\it \Pi }_{\,{\it 2}}$
continues positive, we have that ${\cal M}_{3}$ is imaginary in eq. (4.33).

Finally, we should state that the results obtained in Ref. \cite{Our} and in
this paper for the superconducting properties of the charged anyon fluid
have been derived on the base of a linear approximation. If nonlinear
effects, as for instance vortices, are considered, it is possible that a
richer scenario for the superconducting phases of the anyon system will
appear.

{\bf Acknowledgments}

This research has been supported in part by the National Science Foundation
under Grant No. PHY-9722059.

\end{document}